# 국문 제목

# Optimizing Keyphrase Ranking for Relevance and Diversity Using Submodular Function Optimization (SFO)

Muhammad Umair[a], Syed Jalaluddin Hashmi[a], Young-Koo Lee[b]
{umair, jalal.hashmi, yklee}@khu.ac.kr
[a] Department of Computer Science and Engineering, Kyung Hee University
[b] College of Software, Kyung Hee University

### Abstract

Keyphrase ranking plays a crucial role in information retrieval and summarization by indexing and retrieving relevant information efficiently. Advances in natural language processing, especially large language models (LLMs), have improved keyphrase extraction and ranking. However, traditional methods often overlook diversity, resulting in redundant keyphrases. We propose a novel approach using Submodular Function Optimization (SFO) to balance relevance and diversity in keyphrase ranking. By framing the task as submodular maximization, our method selects diverse and representative keyphrases. Experiments on benchmark datasets show that our approach outperforms existing methods in both relevance and diversity metrics, achieving SOTA performance in execution time. Our code is available [online](online).

## 1. Introduction

Keyphrase ranking is crucial in information retrieval, summarization, and indexing. It assists in identifying and extracting phrases that concisely capture the core concepts of a document. Traditionally, keyphrase extraction methods [1] have focused on selecting highly relevant phrases to the document's content. However, a narrow focus on relevance often results in redundant keyphrases, which fail to capture the full breadth of the document's topics. A recent survey highlights the diversity issue in Keyphrases [2]. Ensuring diversity in keyphrase selection is thus vital to provide a more comprehensive representation of a document's content, particularly in large-scale databases and information retrieval systems.

Recent advancements in natural language processing have been particularly notable in developing large language models such as BERT and MiniLM[1], which have significantly improved the understanding of contextual relationships within text. These models enable more nuanced keyphrase extraction by leveraging deep contextual embeddings. However, even with LLMs, achieving a balance between relevance and diversity in keyphrase selection remains challenging.

Many approaches focus on maximizing relevance or incorporating diversity through heuristic methods, but few offer a theoretically grounded framework that optimally combines both objectives.

To overcome this limitation, we introduce a novel keyphrase ranking method that leverages SFO to balance relevance and diversity. SFO, which exhibit a natural property of diminishing returns, are well-suited for optimization problems where adding new elements to a set yields decreasing incremental benefits. SFO has been successfully applied in various domains requiring diversity, such as document summarization [3] and data subset selection. Despite its potential, the application of submodular optimization to keyphrase extraction remains underexplored. By framing the keyphrase selection task as a submodular maximization problem, our method ensures that the final set of ranked keyphrases is representative and diverse. Our approach integrates SOTA sentence embeddings from pre-trained language models like MiniLM with the

---

[1] http://surl.li/nyhzbz

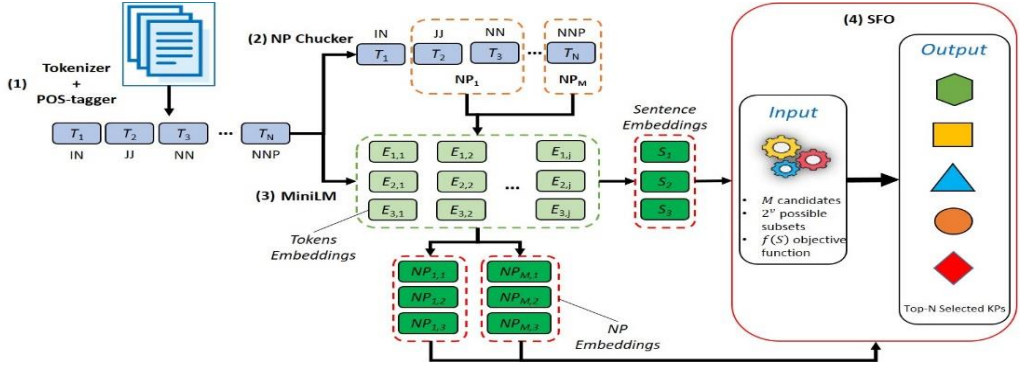
Figure 1: Proposed Architecture

submodular optimization framework. This allows us to capture both the semantic meaning and contextual relevance of keyphrases within the document while ensuring the selected keyphrases cover diverse aspects of the content. Our extensive testing on popular benchmark datasets has shown that our approach surpasses conventional ranking methods in relevance and diversity measures. Additionally, it enhances execution time and scalability, especially for long documents. Our contributions are threefold.

1. We propose a novel method for keyphrase ranking that leverages SFO to optimize the relevance and diversity trade-off.
2. We empirically evaluated our method using benchmark datasets, demonstrating its performance in relevance, diversity metrics, and execution time compared to three baseline techniques.
3. We introduce diversity analysis using Intra-List Distance (ILD) and Subtopic Recall (SR), a novel contribution to the keyphrase extraction field, as previous works have primarily focused on precision, recall, and F1-Score alone.

The rest of this paper is organized as follows: Section 2 covers related work. Section 3, methodology, consists of preliminary, and proposed method. Section 4 discusses experiments and results, with experimental setup, results, and complexity analysis. Finally, Section 5 concludes with a summary and future directions.

## 2. Related Work

Keyphrase extraction is crucial in NLP, serving to condense content and enhance information retrieval systems. While traditional methods like TF-IDF, and TextRank have been instrumental, they often struggle to capture intricate semantic relationships. The advent of machine learning-based models, such as KEA: Practical Automatic Keyphrase Extraction and clustering techniques, has significantly improved keyphrase extraction by leveraging annotated data and document structure. The recent emergence of large language models (LLMs) like BERT and GPT-2 has further advanced keyphrase extraction, particularly in capturing contextual semantics, but redundancy issues persist.

Diversity in keyphrase extraction has been recognized as essential for comprehensive document representation. Techniques like Maximum Marginal Relevance (MMR) and Determinantal Point Process (DPP) have been employed to balance relevance and redundancy in information retrieval and summarization. In keyphrase extraction, methods such as MMR and DPP have improved diversity in EmbedRank++ [4] and [5], respectively. However, these methods often rely on heuristics and lack a formal optimization framework. The underexplored use of SFO in keyphrase extraction, despite its success in domains like document summarization and data selection, presents an opportunity for a principled way to balance relevance and diversity.

## 3. Methodology

In this section, we present preliminary, which introduces the definitions of SFO, and proposed method, describing our keyphrase extraction approach using SFO.

### 3.1 Preliminary

Let $V = \{kp_i\}_{i=1}^{M}$ be the set of all candidate keyphrases, where $kp_i$ represents the $i-th$ candidate keyphrase. The efficacy of a set of keyphrases is modeled by a submodular function $f: 2^V \to \mathbb{R}$, where $2^V$ denotes the power set of $V$, i.e., all subsets of $V$.

For two sets $A \subseteq B \subseteq V$ and an element $x \in V \setminus B$, the function $f$ satisfies the diminishing returns property:
$$f(A \cup \{x\}) - f(A) \geq f(B \cup \{x\}) - f(B),$$
where $A$ and $B$ represent subsets of selected keyphrases and $x$ is a keyphrase not yet included in the set $B$.

The goal of the SFO-based keyphrase ranking is to maximize the function $f(S)$ by selecting a subset $S \subseteq V$ of top-N keyphrases that optimally balances relevance and diversity at the same time.

### 3.2 Proposed Method

The architecture of the proposed method is presented in Figure 1. In this method, we followed the common process of keyphrase extraction and ranking. Begins by tokenizing the input document and tagging each token with its corresponding Part-of-Speech (POS) label. From these tagged tokens, noun phrases (NPs) are extracted, serve as candidate keyphrases. To compare and rank these candidate keyphrases, a pre-trained sentence embedding model, such as MiniLM, is used to generate embeddings for both the NPs and the document. These embeddings are projected into the same dimensional space to facilitate the comparison of relevance between the keyphrases and the document, in our case we chose the size of embedding to 512.

Our objective is to efficiently select a subset $S$ of keyphrases from the candidate set $M$, which represents the set of all candidate keyphrases, to maximizes an objective function $f(S)$ that balances both relevance and diversity at the same time. The objective function $f(S)$ is defined as:

$$f(S) = \sum_{kp \in S} R(kp) - \alpha \sum_{kp_i \neq kp_j} Sim(kp_i, kp_j) \quad (1)$$

Where $R(kp)$ is the relevance score of a candidate keyphrase $kp$, and $Sim(kp_i, kp_j)$ represents the similarity between keyphrase $kp_i$ and $kp_j$. The hyperparameter $\alpha \geq 0$ controls the trade-off between relevance and diversity. The relevance score $R(kp)$ measures how well the keyphrase $kp$ represents the document's content. It is computed using the cosine similarity between the embedding of the keyphrase $e_{kp}$ and the document $e_D$, as defined below:

$$R(kp) = cos(e_{kp}, e_D) = \frac{e_{kp}^T e_D}{||e_{kp}|| \, ||e_D||} \quad (2)$$

The diversity term penalizes the selection of similar keyphrases to promote a diverse set. The similarity between two keyphrases $kp_i$ and $kp_j$ is also measured by cosine similarity.

$$Sim(kp_i, kp_j) = cos(e_{kp_i}, e_{kp_j}) = \frac{e_{kp_i}^T e_{kp_j}}{||e_{kp_i}|| \, ||e_{kp_j}||} \quad (3)$$

Finally, a greedy algorithm is employed to rank and select the top $N$ keyphrases. This algorithm progressively constructs the set $S$ of selected keyphrases by adding one keyphrase at a time, consistently selecting the locally optimal choice that yields the highest marginal gain in the objective function.

Table 1: Dataset statistics: 'GKP' Average Gold KP, 'KPL' Average KP Length, and 'DL' the Average Document Length.

| Sr. | Dataset | GKP | KPL | DL |
|---|---|---|---|---|
| 1 | Inspec | 9.83 | 2.33 | 121.82 |
| 2 | NUS | 11.66 | 2.13 | 7103.20 |
| 3 | SemEval | 15.06 | 2.15 | 7928.28 |

The SFO framework ensures that the selected keyphrases are not only relevant to the document but also diverse, minimizing redundancy in the final selection.

### 4. Experiments, and Results

In this section, we present experimental setup and results, where we evaluate our approach against baseline methods, followed by complexity analysis, which provides a detailed evaluation of the computational complexity of our method.

### 4.1 Experimental Setup

We evaluate our model on three public datasets, with dataset statistics provided in Table 1. Our model is compared against three baseline methods, all of which utilize identical tools for tokenization, POS tagging, and noun phrase chunking under the same experimental setup. The NLTK toolkit is used for tokenization and POS tagging. We assess performance using Precision@N, Recall@N, F1-Score@N, diversity metrics, and execution time, where $N$ equals 5. Additionally, we analyze the impact of the trade-off parameter $\alpha$ on balancing relevance and diversity in comparison to the baseline methods. The overall performance results are summarized in Table 2. Due to space constraints, we have presented only the F1-Score, omitting Precision@N and Recall@N.

Table 2: Overall Performance (F1-Score@5)

| Method | Inspec | NUS | SemEval |
|---|---|---|---|
| SFO | 19.29 | 20.98 | 16.90 |
| EmbedRank++ | 61.85 | 74.18 | 77.05 |
| SIFRank | 65.00 | 66.53 | 66.66 |
| DPP | 62.43 | 31.95 | 23.82 |

### 4.2 Results

In Table 3, varying the trade-off parameter $\alpha$ allowed control over this balance at $\alpha = 0.1$. Relevance was prioritized with higher precision but with some redundancy. In contrast, $\alpha = 0.5$ achieved an optimal F1-Score by balancing relevance and diversity, and $\alpha = 0.9$ emphasized diversity but reduced precision.

Table 3: F1-Score@5 for Different Values of α

| α | Inspec | NUS | SemEval |
|---|---|---|---|
| 0.1 | 17.66 | 25.11 | 18.75 |
| 0.5 | 19.29 | 20.98 | 16.90 |
| 0.9 | 17.59 | 16.10 | 12.48 |

While our method outperformed in execution time in Figure 2 compared to the baselines. DPP is more computationally expensive, especially on the larger NUS and SemEval datasets, taking up to 80.9 seconds per document, our SFO method maintains competitive runtimes across all datasets, achieving a balance between speed and keyphrase extraction quality.

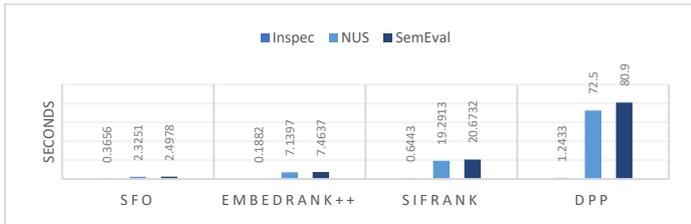

Figure 2: Execution time Comparison

To evaluate the diversity of the keyphrases extracted by our method, we computed the Intra-List Distance (ILD), and Subtopic Recall (SR), on the three datasets.

Our method achieved the highest ILD of 0.8277 presented in Table 4, indicating that the selected keyphrases are more dissimilar to each other compared to those from baseline methods. The Subtopic Recall 0.8144 was also superior, suggesting that our method covers a broader range of topics within the documents.

Table 4: Diversity Metrics Comparison

| Method | ILD | SR |
|---|---|---|
| SFO | 0.8277 | 0.8144 |
| EmbedRank++ | 0.5633 | 0.7326 |
| SIFRank | 0.3464 | 0.6807 |
| DPP | 0.2128 | 0.3734 |

These results confirm that incorporating SFO enhances keyphrase diversity without compromising relevance, as evidenced by the strong performance in both diversity and relevance metrics.

### 4.3 Complexity Analysis

The tokenization, POS tagging, and noun phrase chunking operate in linear time $O(M)$, where $M$ is the number of candidate keyphrases. Embedding generation and relevance score calculation have a complexity of $O(M \cdot d)$, while pairwise diversity scoring adds $O(M \cdot |S| \cdot d)$, where $|S|$ is the selected keyphrase set size. The greedy optimization, selecting keyphrases iteratively, contributes $O(M \cdot N \cdot d)$ per iteration, leading to an overall complexity of $O(M \cdot N^2 \cdot d)$ across $N$ iterations. Hence, the total complexity is $O(M \cdot (1 + N^2) \cdot d)$, which remains efficient as $N$ is small relative to $M$.

### 5. Conclusion and future work

Our method balances relevance and diversity by treating keyphrase ranking as a submodular optimization problem and applying a greedy algorithm. This approach ensures comprehensive, non-repetitive keyphrase selection with improved diversity and efficiency. In the future, we plan to explore adaptive strategies for dynamically adjusting the trade-off parameter based on document characteristics to further optimize this balance across various datasets.

### Acknowledgment

This work was supported by Institute of Information & communications Technology Planning & Evaluation (IITP) grant funded by the Korea government(MSIT) (No.RS-2022-00155911, Artificial Intelligence Convergence Innovation Human Resources Development (Kyung Hee University)).